\documentclass[a4paper]{article}
\usepackage{anysize}
\marginsize{2cm}{2cm}{2cm}{2cm}

\usepackage{graphicx}
\usepackage{physics}

\usepackage{tikz}
\usepackage{xargs,ifthen,xstring,xparse,etoolbox,mathtools}
\usepackage{environ}
\usetikzlibrary{quantikz}

\usepackage{multirow}

\usepackage[left=1.5cm,right=1.5cm,top=1.5cm,bottom=1.5cm,ignoreheadfoot]{geometry}
\usepackage{array}
\usepackage{hyperref}
\usepackage{amsmath}
\usepackage{amsfonts}
\usepackage{dsfont}
\usepackage{bbm}
\usepackage{subcaption}

\usepackage{authblk}

\usepackage{csquotes}

\usepackage[numbers]{natbib}

\NewDocumentCommand\citelong{m O{}}{%
    \citeauthor{#1}~(\citeyear{#1})#2\cite{#1}%
}

\title{Quantum Poker\\A game for quantum computers suitable for benchmarking error mitigation techniques on NISQ devices}

\author[$\dagger$]{Franz G. Fuchs}
\author[$\ddagger$]{Vemund Falch}
\author[$\ddagger$]{Christian Johnsen}
\affil[$\dagger$]{SINTEF AS, Department of Mathematics and Cybernetics, Oslo, Norway}
\affil[$\ddagger$]{NTNU, Department of Physics, Trondheim, Norway}

\date{\today}

\begin{document}

\maketitle
\begin{abstract}
Quantum computers are on the verge of becoming a commercially available reality.
They represent a paradigm shift in computing, with a steep learning gradient.
The creation of games is a way to ease the transition for beginners.
We present a game similar to the Poker variant Texas hold 'em with the intention to serve as an engaging pedagogical tool to learn the basics rules of quantum computing.
The concepts of quantum states, quantum operations and measurement can be learned in a playful manner.
The difference to the classical variant is that the community cards are replaced by a quantum register that is "randomly" initialized, and the cards for each player are replaced by quantum gates, randomly drawn from a set of available gates.
Each player can create a quantum circuit with their cards, with the aim to maximize the number of $1$'s that are measured in the computational basis.
The basic concepts of superposition, entanglement and quantum gates are employed.
We provide a proof-of-concept implementation using Qiskit\cite{Qiskit}.
A comparison of the results for the created circuits using a simulator and IBM machines is conducted, showing that error rates on contemporary quantum computers are still very high.
For the success of noisy intermediate scale quantum (NISQ) computers, improvements on the error rates and error mitigation techniques are necessary, even for simple circuits.
We show that quantum error mitigation (QEM) techniques can be used to improve expectation values of observables on real quantum devices.
\end{abstract}

\section{Introduction}

Quantum computing is an emerging technology exploiting quantum mechanical phenomena -- namely superposition, entanglement, and tunneling -- in order to perform computation.
Quantum computers have huge potential to transform society in a similar way that classical computers have, because they open up the possibility to tackle certain types of problems that are beyond the reach of classical computers.
The first commercially available quantum computers are expected within the next five years, and
it is expected that quantum computers will outperform their classical counterparts in some tasks within the same time period. In order to utilize the potential power of quantum computers, one has to formulate a given problem in a form that is suitable for a quantum computer (encoding step) and develop specialized algorithms. These type of algorithms are fundamentally different from classical algorithms.
Getting accustomed to quantum algorithms has a considerable learning curve and requires a multidisciplinary approach. Typically, knowledge from physics, mathematics, computers science and a firm understanding from an application area such as quantum chemistry, optimization, or machine learning is required.
In addition, it is advantageous to have knowledge of the underlying physics, particularly in the NISQ era.

The New York Times estimated in October 2018\footnote{https://www.nytimes.com/2018/10/21/technology/quantum-computing-jobs-immigration-visas.html} that the global number of high-level researchers in quantum computing may be less than a thousand.
The design of games that make use of the underlying rules of quantum computers is a way to attract more interest and ease the transition from classical algorithms to quantum algorithms for beginners. 
In order to play the basic version of this game, knowledge of quantum physics or advanced mathematics is not required.
Just as with any other game, one has to learn a set of rules, which are described in Section~\ref{sec:QPRules}.
Since one of the best ways to learn is through play, this game can help to attract more people -- from elementary school to post-graduate level at the university -- to the field of quantum computing and we hope that the game awakens curiosity for present challenges in the field, such as the design of quantum algorithms and error mitigation schemes.

Near-term applications of early quantum devices, such as electronic structure problems and optimization, rely on accurate estimates of expectation values to become of practical relevance.
However, inherent noise in quantum devices leads to wrong estimates of the expectation values of observables.
Therefore, getting rid of (most of) the noise inherent in quantum computing is a critical step toward making it useful for practical applications. 
\textit{Quantum error correction (QEC)} can only be achieved by increasing quantum resources (ancillary qubits). The first scheme was proposed by \citelong{Shor1995} and many other schemes were proposed since then, e.g., the class called stabilizer codes, see \citelong{gottesman1997stabilizer}.
However, the number of ancillary qubits needed to achieve QEC depends intrinsically on the error rates and is out of reach for NISQ devices.
\textit{Quantum error mitigation (QEM)}, on the other hand can be achieved with additional classical resources only and is therefore applicable to NISQ devices.

The main contributions of this article are as follows.
\begin{itemize}
    \item We introduce a novel quantum game based on classic Poker. The game is useful to introduce basic quantum computing concepts to beginners.
    \item We implement error mitigation schemes based on the extrapolation technique. 
    A comparison of the results on simulators and real quantum devices is provided.
    We show that errors can successfully be mitigated when one is interested in expectation values of an observable.
\end{itemize}

This article is organized as follows. After describing related work in Section~\ref{sec:relatedwork} we present the description of our game in Section~\ref{sec:QPRules}.
Using a representative circuit from an example game we report results of ideal and noisy circuit sampling in Section~\ref{sec:sim}.
Finally, we describe methods for error mitigation in Section~\ref{sec:errormitigation} before concluding in Section~\ref{sec:conclusion}.

\section{Related Work}\label{sec:relatedwork}
There exist a number of games based on quantum physics and they can be categorized into the following two types.
The first type attempt to illustrate quantum mechanical effects, and one might therefore call them quantum mechanics games.
The second type illustrate quantum computing via qubits and quantum circuit building.
It is the latter type of game that has been developed in connection with this paper. 
For a recent review of the subject of quantum games we refer to, e.g., \cite{khan2018quantum}.
Existing quantum computing games include "Battleships with partial NOT gates", solving puzzles by creating simple programs, and "quantum chess". Some are available as Jupyter Notebooks in Qiskit's Github repository of tutorials\footnote{Qiskit tutorials - Quantum games \url{https://github.com/Qiskit/qiskit-tutorials-community/tree/master/games}}.

Mitigating the effect of noise on the execution of circuits is critical for the success of NISQ devices. The ideal action of a gate is given by a unitary operator $U$ transforming a state $\ket{\phi}$ into $U\ket{\phi}$. The are two basic types of noise.
\textit{Coherent noise} means that a small perturbation $\widetilde{U}$ of $U$ is executed, where $\widetilde{U}$ is still unitary and preserves the purity of the state $\ket{\phi}$. An example is a slight over-rotation.
\textit{Incoherent noise} does not preserve the purity of the state. This type of noise comes from the (unwanted) interaction with the environment.
In this case the evolution must be described through density matrices and Kraus operators.
An example of incoherent noise is amplitude damping modeling relaxation from an excited state to the ground state. For a single qubit with decay probability $p$, the density matrix $\rho = \ket{\phi}\bra{\phi}$ is mapped to $K_0 \rho K_0^\dagger + K_1 \rho K_1^\dagger$ with $K_0=\begin{pmatrix}
    1 & 0\\
    0 & \sqrt{1-p} 
    \end{pmatrix}, K_1=\begin{pmatrix}
    0 & \sqrt{p}\\
    0 & 0 
    \end{pmatrix}$.
Different types of techniques have been presented in the literature that can be used to mitigate the influence of noise on the ideal circuit. In the following we discuss the most important ones.

\emph{Quantum subspace expansion (QSE).}
The main idea is to use operators from a given set (such as the set of Pauli operators) to expand about the variational solution. A generalized eigenvalue problem using linear subspaces is solved.
This method leaves the circuit width (number of qubits) and depth (largest number of gates on any input–output path) unchanged.
The method is introduced by \citelong{Wecker2015} where numerical evidence is provided. For the variational quantum eigensolver numerical evidence for error mitigation is provided by \citelong{McClean2017}[ in ].
An extension of the approach is given in \citelong{Colless2018}. The complete energy spectrum of the H2 molecule is calculated with near chemical accuracy.
\citelong{premakumar2018error} generalize the concept of decoherence-free subspace (DFS) to noise with correlations in space to avoid regions where decoherence measures are high. However, as the authors point out themselves, only dephasing noise is considered and it is unlikely that the method is useful for general error models. DFSs do not exist for other types of noise, e.g., if noise flips spins.

\emph{Probabilistic error cancellation.}
The main idea is to represent the ideal circuit as a quasi-probabilistic mixture of noisy ones.
The circuit depth and width remain unchanged with this method.
\citelong{Temme2017} present the method together with numerical evidence.
\citelong{Song2019} demonstrate an error mitigation protocol based on gate set tomography and quasiprobability decomposition.  One- and two-qubit circuits are tested on a superconducting device, and computation errors are successfully suppressed.
Process tomography is not feasible for more than a few qubits since it scales exponentially with the number of qubits.
In addition, process tomography is sensitive to noise in the pre- and post rotation gates plus the measurements (SPAM errors). Gate set tomography can take these errors into account, but the scaling becomes even worse.

\emph{Extrapolation techniques.}
The main idea is to amplify the noise deliberately in a controlled way. The information of the dependence of the expectation value on the noise level is used to extrapolate back to the zero noise level. The circuit width remains unchanged, but the circuit depth is longer (or gate times are prolonged in case of phase control).
\citelong{Temme2017} and \citelong{Li2017} introduced the technique and provide numerical evidence.
\citelong{Endo2018} extend the work of \cite{Temme2017,Li2017} by accounting for the inevitable imperfections in the experimentalist's measuring the effect of errors in order to design efficient QEM circuits.
\citelong{Kandala2019}
presents important considerations for hardware and algorithmic implementations of the zero-noise extrapolation technique, and demonstrates tremendous improvements in the accuracy of variational eigensolvers implemented by a noisy superconducting quantum processor. 
Evidence on real quantum hardware is presented. In contrast to previous works, the increase of errors is not done by artificially introducing additional gates, but directly by pulse control. For IBM's quantum computers, access at this level (pulse/machine level) is only possible for customers of, which makes this technique inaccessible to us.
\citelong{Otten2019}
introduce a technique that can be viewed as a multidimensional extension.
The approach corresponds to repeating the same quantum evolution many times with known variations on the underlying systems' error properties.
They show that the effective spontaneous emission, T1, and dephasing, T2, times can be increased using their method in both simulation and experiments on an actual quantum computer.

In this article we present results using the extrapolation method.
We conduct experiments on simulators with error models and real quantum computers.

\begin{figure}
    \centering
    \begin{tabular}{m{10cm} m{10cm}}
    \begin{subfigure}[b]{0.55\textwidth}
    \includegraphics[width=\textwidth]{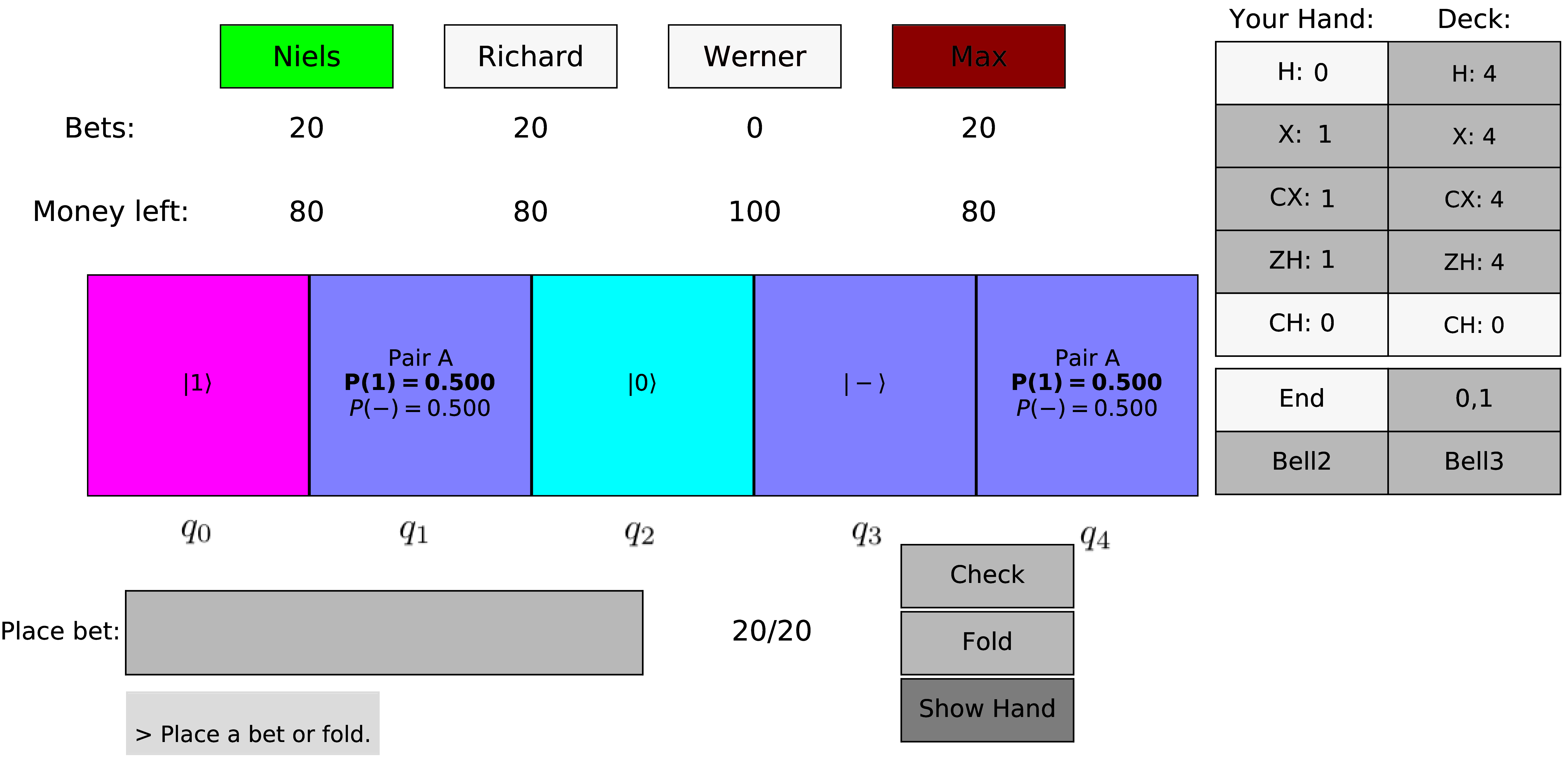}
        \caption{\sf A screenshot from the game at the end of all betting rounds.
        Niels (green) is the dealer and Max (red) is revealing his cards.
        For the community cards the color cyan, blue, pink indicate a 0\%, 50\%, 100\% probability to obtain a 1, respectively.
        }
        \label{fig:screenshot_game}
    \end{subfigure}
    &
    \vspace{1\baselineskip}
    \begin{subfigure}[b]{.4\textwidth}
        \resizebox{1\hsize}{!}{
            \begin{tikzcd}
                \lstick{ $q_{0} : \ket{0}$ } & \gate{X}\gategroup[5,steps=6,style={dashed,rounded corners,fill=blue!20, inner xsep=2pt},background]{ "community cards"} & \ctrl{2} & \qw & \qw & \ctrl{4} & \qw %\barrier[0em]{4}
                & \qw & \qw & \qw & \qw %\barrier[0em]{4}
                & \qw \\%& \meter & \qw & \qw & \qw & \qw & \qw \\
                \lstick{ $q_{1} : \ket{0}$ } & \gate{X} & \qw & \ctrl{1} & \targ{} & \qw & \qw & \qw & \qw & \qw & \qw & \qw \\%& \qw & \meter & \qw & \qw & \qw & \qw \\
                \lstick{ $q_{2} : \ket{0}$ } & \qw & \targ{} & \targ{} & \qw & \qw & \ctrl{2} & \qw & \gate{X}\gategroup[2,steps=3,style={dashed,rounded corners,fill=red!20, inner xsep=2pt},background]{"Max's cards"} & \qw & \ctrl{1} & \qw \\%& \qw & \qw & \meter & \qw & \qw & \qw \\
                \lstick{ $q_{3} : \ket{0}$ } & \gate{H} & \gate{Z} & \qw & \qw & \qw & \qw & \qw & \gate{Z} & \gate{H} & \targ{} & \qw \\%& \qw & \qw & \qw & \meter & \qw & \qw \\
                \lstick{ $q_{4} : \ket{0}$ } & \gate{H} & \gate{Z} & \qw & \ctrl{-3} & \targ{} & \targ{} & \qw & \qw & \qw & \qw & \qw %& \qw & \qw & \qw & \qw & \meter & \qw \\
            \end{tikzcd}
        }
    \vspace{1.675\baselineskip}
        \caption{\sf Circuit that Max creates by placing his X,\\ ZH and CX gates on the community cards. By doing so he changes the quantum state to obtain more 1's with a higher probability.}
        \label{fig:Maxscircuit}
    \end{subfigure}
    \end{tabular}

    \vspace{1\baselineskip}    
    \begin{subfigure}[b]{1\textwidth}
        \centering
        \begin{tabular}{ l l l l l l }
        $X\ket{0}=\ket{1}$ & $H\ket{0}=\ket{+}$ & $Z\ket{0}=\ket{0}$ & $CX\ket{00}=\ket{00}$ &  $CX\ket{+0}=\ket{00}+\ket{11}$ & $CX\ket{++}=\ket{++}$\\
        $X\ket{1}=\ket{0}$ & $H\ket{1}=\ket{-}$ & $Z\ket{1}=\ket{1}$ & $CX\ket{01}=\ket{01}$ & $CX\ket{+1}=\ket{01}+\ket{10}$ & $CX\ket{+-}=\ket{--}$\\
        $X\ket{+}=\ket{+}$ & $H\ket{+}=\ket{0}$ & $Z\ket{+}=\ket{-}$ & $CX\ket{10}=\ket{11}$ & $CX\ket{-0}=\ket{00}-\ket{11}$ & $CX\ket{-+}=\ket{-+}$\\
        $X\ket{-}=\ket{-}$ & $H\ket{-}=\ket{1}$ & $Z\ket{-}=\ket{+}$ & $CX\ket{11}=\ket{10}$ & $CX\ket{-1}=\ket{01}-\ket{10}$ & $CX\ket{--}=\ket{+-}$
        \end{tabular}
        \caption{\sf List of useful rules showing the effect of gates on quantum states.
        The symbols $X, H, Z, CX$ represent the available cards and $\ket{.}$ is the state of the quantum register they act on.
        Global phase and normalization have been ignored.
        }
        \label{tab:Gates}
    \end{subfigure}

    \caption{
        Example of a quantum Poker game and the basic rules needed to play.
        The state of the quantum register before each player places their gates, is shown in the middle of (a).
        Max has one out of total four X, CX, and ZH gates and no H gate.
        (b) shows the underlying circuit generating the quantum register.
        Note that qubit $q_2$ is entangled with qubits $q_0$ and $q_1$, but the resulting state is shown to be $\ket{0}$ in (a), since the bit gets flipped twice.
        In order to win, Max needs to alter the state of the "community cards" in such a way that the states of the qubits get "closer to $\ket{1}$.
        Looking at the list of operations shown in (c) Max applies his X-gate to change the state of $q_2$ to be $\ket{1}$.
        He then applies the ZH gate to qubit $q_3$ to obtain the state $\ket{0}$ and flips it to $\ket{1}$ by applying a CX with qubit $q_2$, which is in state $\ket{1}$.
    }
    \label{fig:examplegame}
\end{figure}

\section{Quantum Poker Rules}\label{sec:QPRules}
The classical Texas hold'em round involves five community cards on the table shared by all the players, while each player holds 2 unique cards in their own hand. The small and big blind bets are placed at the start of each round, relative to a rotating dealer position. The community cards are gradually revealed, the first three - called "the flop" - are revealed simultaneously, and the last two called "the turn" and "the river" are revealed one at a time.
The players take turns betting at the start of each round. In order to stay in the game one has to match or increase the highest bet currently in play. Otherwise, one can fold and forfeit the chance to win the pot, i.e., the sum of money that players wager during a single hand or game.
After all five cards have been revealed there is a final round of betting before the players can choose to either show their hand in the hope of winning or folding it and forfeiting the pot. The goal is to combine up to five cards from both your own hand and the table to form the best Poker hand
(hands are ordered by their probability and for equal probability by the "highest" card)
out of all the players. The player or players with the best hand win(s) the pot.

The quantum Poker game considered here draws inspiration from Texas hold 'em Poker and shares its structure.
The betting rounds in the two games are identical. 
Community cards are replaced with qubits and the cards received by each player are replaced by quantum logic gates which can be applied to the qubits, i.e., community cards.
\begin{itemize}
    \item 
In quantum Poker each player acts on a personal "copy" of the \textit{community cards}, consisting of a quantum register.
Hence, the only interaction between the players is through betting.
An example initial state is shown in Figure~\ref{fig:examplegame}.
The qubits can be in an any state, e.g. the ground state $\ket{0}$, the excited state $\ket{1}$, a state of superposition $\ket{\pm} = \frac{1}{\sqrt{2}}(\ket{0} \pm \ket{1})$, or in an entangled state like the Bell states $\frac{1}{\sqrt{2}}(\ket{00} \pm \ket{11})$.
\item
The \textit{personal cards} come from a set of available quantum logic gates which can be applied to the qubits.
This set is known to all players, such that they can deduce the probability of another player having, e.g., a CX gate.
The set of quantum logic gates consist of operations acting on one or two qubits.
In our implementation, the set consists of the Hadamard gate H, the phase flip gate Z, the NOT gate X and the controlled NOT gate CX.
The action of each gate applied to the $(\ket{1},\ket{0})$- and $(\ket{+},\ket{-})$-basis is shown in Figure~\ref{tab:Gates}.
\item
The \textit{final score} of each player is the number of qubits measured to the state $\ket{1}$ in the computational basis at the end of the round.
The winner(s) of each pot is/are the player(s) with the highest score.
\end{itemize}

The game is designed such that it can be played without knowledge of physics and advanced mathematics. The only rules to learn are given in the list shown in Figure~\ref{tab:Gates} and the following rules: The state $\ket{1}$ has a 100\% probability to "give a 1", $\ket{0}$ a 0\% probability to "give a 1", and the states $\ket{+}, \ket{-}$ a 50\% chance to "give a 1".
In addition, states marked as "Pair" the probabilities of the two qubits cannot be described independently. As an example, the Bell states $\frac{1}{\sqrt{2}}(\ket{00} \pm \ket{11})$ "give two 1s" with a probability of 50\%.
For the sake of correctness, the expression "give a 1/give two 1s" refers to measurement in the computational basis.
The idea of the game is that players will get familiar with the rules underlying quantum computing.

In the quantum Poker game a state
\begin{equation}\label{eq:QP}
    \ket{\phi} = VU \ket{0}^{\otimes n}
\end{equation}
is created.
The matrix $U$ creates the $n$ "community cards" and is equal for all players.
The matrix $V$ is created by the players  through placing their "personal cards".
Given the state $\ket{\phi}$ and an observable $A$, the expectation value of $A$ in the state $\phi$ is given by
\begin{equation}
    \langle A \rangle_\phi \coloneqq \bra{\phi} A \ket{\phi} = \sum_i \lambda_i \left| \langle\phi|\psi_i\rangle\right|^2.
\end{equation}
Here, A is a self-adjoint operator on the Hilbert space $\mathbb{C}^{\otimes n}$, and $\{\lambda_i,\ket{\psi_i}\}$ is the set of eigenvalues and eigenvectors of $A$.
To match the objective of our game, we need to define an observable $A$ such that $\langle A \rangle_\phi$ is equal to the expected number of ones in the computational basis.
This can be done by choosing
\begin{equation}\label{eq:numberoperator}
    A = \sum_{i=1}^{2^5} b(i) P_i,
\end{equation}
where $b(i)$ is a function returning the number of ones of the binary representation of $i$,
and $P_i = \ket{i}\bra{i}$ is the measurement operator in the computational basis.
A is a diagonal matrix with eigensystem $\{b(i),\ket{i}\}$.
The matrix $A$ can also be constructed  via the number operator in the second quantization (a formalism used to describe and analyze quantum many-body systems), which is given by
\begin{equation}
    A = \sum_i N_i, \quad \text{ where } N_i = a_i^\dagger a_i.
\end{equation}
The creation and annihilation operators are given by
\begin{equation}
    \begin{split}
        a_i^\dagger &= I^{\otimes n-i-1} \otimes Q^+ \otimes \sigma_z^{\otimes i}, \\
        a_i &= I^{\otimes n-i-1} \otimes Q^- \otimes \sigma_z^{\otimes i}, \\
    \end{split}
\end{equation}
and the raising and lowering operator is given by
\begin{equation}
    Q^\pm = \frac{1}{2} \left( \sigma_x \mp i \sigma_y \right),
\end{equation}
i.e, $Q^+=\begin{pmatrix}
    0 & 1\\
    0 & 0 
    \end{pmatrix},
    Q^-=\begin{pmatrix}
    0 & 0\\
    1 & 0 
    \end{pmatrix}$.
As an example, for two qubits $A$ is a diagonal matrix with entries $(0,1,1,2)$, from upper left to lower right.

\begin{figure}
    \centering 
    \begin{tabular}[t]{lr}
    \begin{subfigure}[b]{0.45\textwidth}
        \includegraphics[width=1.\textwidth]{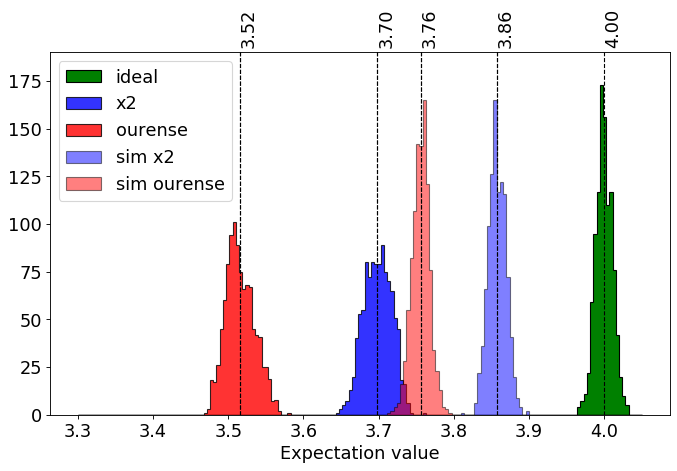}
        \caption{Distribution of the expectation value for Max's circuit.}
    \end{subfigure}
&
    \begin{subfigure}[b]{0.45\textwidth}
        \includegraphics[width=1\textwidth]{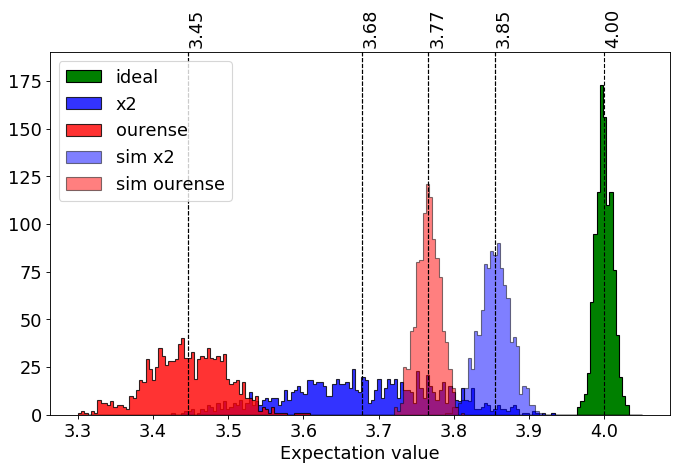}
        \caption{Same as (a), but with Pauli-twirling, see Section~\ref{sec:Paulitwirling}.}
        \label{fig:distributionwithpauli}
    \end{subfigure}
    \\
    &\\
       \begin{subfigure}[b]{0.45\textwidth}
        \includegraphics[width=1\textwidth]{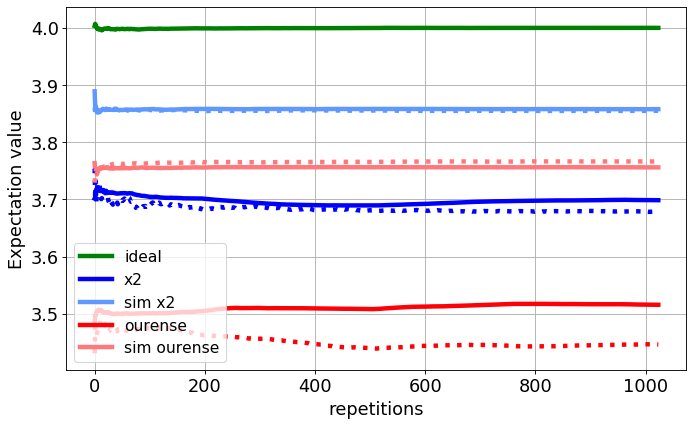}
        \caption{Convergence of the expectation values. Dashed lines represent the Pauli-twirled versions of the experiments.}
        \label{fig:convergenceE}
    \end{subfigure} 
        &
        \begin{parbox}{0.46\textwidth}{\hskip -4.5cm \vskip -6.5cm
                {    \caption{Result for Max's circuit shown in Figure~\ref{fig:Maxscircuit} using simulators and IBM's quantum devices. In total 1024 repetitions with 8192 shots each were used.
                The error models are more optimistic than the real quantum devices.
                Quantum noise on a real device leads to convergence of the expectation value to a much lower value than its simulated noise model, which is in turn lower than the theoretical value.
                A comparison of (a) and (b) shows that Pauli-twirling, see Section~\ref{sec:Paulitwirling}, has little to no effect on the obtained expectation value, even when noise is present.
                It leads, however, to an increased variance in the presence of noise.
    }
            \protect 
    \label{fig:Expectationvalues}
            }}
        \end{parbox}
\end{tabular}
\end{figure}
Using Equation~\ref{eq:numberoperator}, it is easy to calculate the expectation value as
\begin{equation}
    \expval{A}{\Phi} = \bra{\Phi} \sum_{i=1}^{2^5} b(i) P_i\ket{\Phi} =
    \sum_{i=1}^{2^5} b(i) \bra{\Phi}P_i\ket{\Phi} = 
    \sum_{i=1}^{2^5} b(i) \bra{\Phi}P^\dagger_i P_i\ket{\Phi} = 
    \sum_{i=1}^{2^5} b(i) p(i),
\end{equation}
see Postulate 3 \cite[page 84]{nielsen2000quantum}, where $p(i)$ is the probability that result $i$ occurs.
In order to get the expectation value, we can therefore measure the states $\ket{\Phi}$ in the computational basis and multiply the resulting bit strings with $b(i)$.

\section{Ideal and noisy quantum circuit sampling}\label{sec:sim}
In the following we use Max's circuit, shown in Figure~\ref{fig:Maxscircuit}, as an example to investigate the effects of noise in NISQ devices.

\subsection{Ideal simulator}
The state that Max creates is given by $\ket{\phi_\text{Max}} = \frac{1}{\sqrt{2}}(\ket{01101} + \ket{11111})$.
Each realization of the circuit on an ideal (simulated) quantum computer results in a classical bit string $q_{n-1} \dots q_1 q_0$ after measurement.
A state $\phi = \sum_i \alpha_i\ket{i}$ induces a probability distribution $P_\phi(i) = |\alpha_i|^2$.
For Max's circuit this distribution is thus given by a 50\% chance of being in either state $\ket{01101}$ and $\ket{11111}$.
The expectation value for Max's circuit is thus $\langle A \rangle_{\ket{\phi_\text{Max}}} = 4$.
Figure~\ref{fig:Expectationvalues} shows the convergence of sequence averages on an ideal simulator with respect to the number of repetitions to the expectation value, as well as the according probability distribution of the sequence.
Here, we have used Qiskit's simulator with 1024 repetitions, each consisting of 8192 shots.

\subsection{Set of universal quantum gates and circuit mappings}
On a real quantum device, such as IBM's QX architectures, only certain types of operations/gates are supported. This set contains the single qubit gates $U_1, U_2, U_3$, and the CX gate.
In order to execute a circuit, one needs to express the circuit in these basis gates.
The Solovay-Kitaev theorem guarantees that this is always possible up to a given accuracy.

\begin{figure}
\begin{center}
    \begin{tabular}{m{.45\textwidth} m{.45\textwidth}}
     \begin{subfigure}[]{0.5\textwidth}
    \includegraphics[width=.6\textwidth]{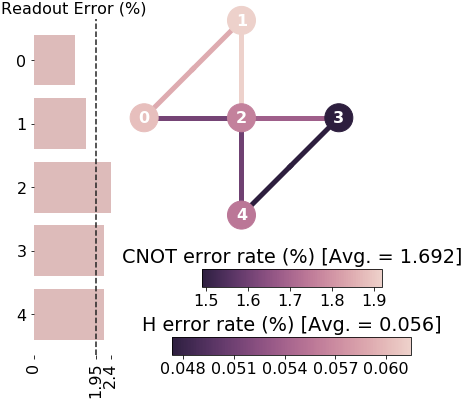}
        \caption{Connectivity and error rates on IBM's QX2.\\ \phantom{ab} (Feb 6 2020)
        }
         \label{fig:ibmqx2}
     \end{subfigure}
        &
    \begin{subfigure}[]{0.5\textwidth}
    \includegraphics[width=.6\textwidth]{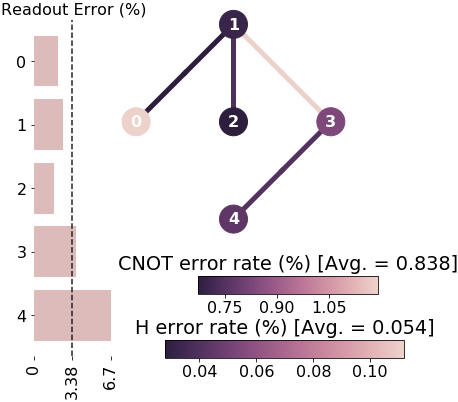}
        \caption{Connectivity and error rates on IBM's ourense.\\ \phantom{ab} (Feb 6 2020)
        }
         \label{fig:ibmourense}
     \end{subfigure}\\
     &\\
    \begin{subfigure}[]{.4\textwidth}
    \resizebox{.7\hsize}{!}{
        \begin{tikzcd}
            \qw & \gate{U_3} & \ctrl{2} & \qw & \qw & \qw & \qw & \qw & \ctrl{2} & \qw & \qw & \qw & \qw  \\
            \qw & \gate{U_3} & \qw & \ctrl{1} & \qw & \qw & \qw & \targ{} & \qw & \qw & \qw & \qw & \qw \\
            \qw & \qw & \targ{} & \targ{} & \ctrl{2} & \targ{} & \ctrl{2} & \ctrl{-1} & \targ{} & \targ{} & \qw & \qw & \qw \\
            \qw & \qw & \qw & \qw & \qw & \qw & \qw & \qw & \qw & \qw & \qw & \targ{} & \qw \\
            \qw & \gate{U_2} & \qw & \qw & \targ{} & \ctrl{-2} & \targ{} & \qw & \qw & \ctrl{-2} & \gate{U_3} & \ctrl{-1} & \qw \\
        \end{tikzcd}
    }
        \caption{Max's circuit transpiled for IBM's QX2.\\
        \phantom{ab} \#CX gates = 9, depth = 11.}
        \label{fig:adapted_circuitqx2}
    \end{subfigure}
    &
    \begin{subfigure}[]{.4\textwidth}
        \resizebox{1.05\hsize}{!}{
        \begin{tikzcd}
            \qw & \gate{U_3} & \qw & \qw & \qw & \qw & \qw & \qw & \qw & \ctrl{1} & \gate{U_3} & \ctrl{1} & \gate{U_3} & \qw & \qw & \qw & \qw & \qw \\
            \qw & \gate{U_2} & \ctrl{1} & \gate{U_3} & \ctrl{1} & \gate{U_3} & \ctrl{2} & \gate{U_3} & \ctrl{2} & \targ{} & \gate{U_3} & \targ{} & \gate{U_2} & \targ{} & \targ{} & \qw & \qw & \qw \\
            \qw & \gate{U_3} & \targ{} & \gate{U_3} & \targ{} & \gate{U_2} & \qw & \qw & \qw & \qw & \qw & \qw & \qw & \ctrl{-1} & \qw & \qw & \qw & \qw \\
            \qw & \gate{U_3} & \qw & \qw & \qw & \qw & \targ{} & \gate{U_3} & \targ{} & \gate{U_3} & \qw & \qw & \qw & \qw & \ctrl{-2} & \gate{U_3} & \ctrl{1} & \qw \\
            \qw & \qw & \qw & \qw & \qw & \qw & \qw & \qw & \qw & \qw & \qw & \qw & \qw & \qw & \qw & \qw & \targ{} & \qw \\
        \end{tikzcd}
        }
        \caption{Max's circuit transpiled for IBM's ourense.\\
        \phantom{ab} \#CX gates = 9, Depth = 16.}
        \label{fig:adapted_circuitourense}
    \end{subfigure}
    \end{tabular}

\end{center}
    \caption{
    A comparison of the two IBM devices used, shows that the qubits have better connectivity on the QX2 device (6 vs 4 bus resonators), the average readout error is much smaller, but the CX error rate is worse, as compared to IBM's ourense.
    In order to execute Max's circuit, shown in Figure~\ref{fig:Maxscircuit}, equivalent circuits, shown in (c) and (d). Both use the same number of CX gates, but have different depth.
    The variables of the $U_2, U_3$ gates are skipped for simplicity.
    The average depth of the circuits can be reduced to a constant after combining neighboring single qubit gates.
    }
    \label{fig:realdevices}
\end{figure}
An additional complication comes from the fact that only a subset of qubits are physically connected.
On IBM's QX devices CX gates can only be applied to qubits that are connected by a bus resonator.
Figures~\ref{fig:realdevices}\subref{fig:ibmqx2},\subref{fig:ibmourense} show the connectivity graph of two of IBM's quantum devices, where edges mark qubits that are physically connected.
In order to execute a circuit with CX gates between two not connected qubits, additional gates, such as SWAP or BRIDGE gates,
need be used to transform the circuit into an equivalent one that obeys the connectivity graph.
Inserting one SWAP or BRIDGE gate increases the number of CX gates by three.
On current NISQ devices the noise level of two-qubit gate (CX) times and error rates are one order of magnitude higher than for single qubit gates\cite{IBMQX2}, see also Figures~\ref{fig:realdevices}\subref{fig:ibmqx2},\subref{fig:ibmourense}.
One therefore wishes to find a mapping with the lowest number of CX gates.
In general, the problem of finding an optimal mapping is $\mathcal{N}\mathcal{P}$-complete problem\cite{Wille2019}.
For recent heuristics we refer to \cite{Itoko2020,Wille2019} and references therein.
For Max's circuit it is easy to find an optimal mapping for IBM's QX2 device manually using only one extra SWAP gate. For the IBM's ourense device we have used the built-in optimizer.
The resulting circuits which are shown in Figures~\ref{fig:realdevices}\subref{fig:adapted_circuitqx2},\subref{fig:adapted_circuitourense} have the same number of CX gates.
However, the circuit adapted to the ourense device has a larger depth due to lower connectivity of the qubits.

\subsection{The effect of noise on quantum computation}
Noise is inherent to quantum computers.
Qiskit provides methods for automatic generation of approximate noise models matching a given hardware device. This enables us to simulate the effects of realistic noise on our computation before we run our circuits on a real quantum computer.
Figure~\ref{fig:Expectationvalues} shows the results of the Max's circuit on different simulators and real quantum devices, using the transpiled circuits shown in Figure~\ref{fig:realdevices}.
Due to the influence of noise, the resulting expectation values converge to a value around $3.86$/$3.76$ for the simulated noise model and $3.70$/$3.52$ on the IBM's quantum computers for the QX2/ourense device.
The gate noise level of IBM's ourense device is slightly better than that of IBM's QX2, but the average readout error and connectivity (and hence on average the depth of a circuit) is worse for the ourense device.
In our case, we achieve better results with the QX2 device, since we can use a shorter depth circuit with less readout errors.
Both values are far off the ideal value of $4$, obtained through a noiseless simulation.
In the following we will show how to mitigate the effect of errors on expectation values in order to get a better estimate of the ideal expectation value.

\section{Error mitigation.}\label{sec:errormitigation}
In this section we will apply the zero-noise extrapolation method to our circuit.
The basic assumption of the method is that the expectation value of an observable depends smoothly on a small noise parameter $\lambda \ll 1$ and admits the following power series,
\begin{equation}
    \langle A \rangle_{\ket{\phi}}(\lambda) = \langle A \rangle_{\ket{\phi}}^* + \sum_{i=1}^n a_i \lambda^i + \mathcal{O}(\lambda^{i+1}),
\end{equation}
where $\langle A \rangle_\phi^*$ is the zero noise value we are trying to recover.
Richardson's deferred approach to the limit\cite{Richardson1927} can then be applied to get a better estimate of the zero noise value.
The method requires to generate $n$ estimates to the expectation value, i.e., $\langle A \rangle_\phi(r_i\lambda)$ for $r_1<r_2<\dots<r_n$.
A better estimate of $\langle A \rangle_\phi^*$ is then constructed by combining these values in such a way that the lowest order terms in the power series cancel.
As an example we can get a second order approximation of the expectation value by combining the results for $r_1=1$ and $r_2=2$ through
\begin{equation}
    2 \langle A \rangle_{\ket{\phi}}(\lambda) - \langle A \rangle_{\ket{\phi}}(2\lambda) = \langle A \rangle_{\ket{\phi}}^* + \mathcal{O}(\lambda^2).
\end{equation}
Clearly, using $r_1=1$ generates the expectation value with the least noise.
Amplification of noise with the factors $r_i>1$ can either be achieved directly through pulse control or through modifying the circuit by adding certain extra gates.
For IBM's QX devices pulse control on devices with more than one qubit is only accessible for their customers, which leaves us with the second possibility.

\begin{figure}
    \centering 
    \begin{subfigure}[b]{0.35\textwidth}
        \begin{tikzcd}
          \qw & \gate{\sigma_a}\gategroup[2,steps=3,style={dashed,rounded corners,fill=blue!20},background]{Pauli-twirling} & \ctrl{1} & \gate{\sigma_c} & \gate{\sigma_e}\gategroup[2,steps=1,style={dashed,rounded corners,fill=red!20, inner xsep=2pt},background,label style={label position=below,anchor=north,yshift=-0.2cm}]{Noise ampl.} & \qw  \\
          \qw & \gate{\sigma_b} &  \targ{} & \gate{\sigma_d} & \gate{\sigma_f} & \qw
        \end{tikzcd}
        \caption{Pauli-twirling and noise amplification.}
        \label{fig:pauliandnoisecircuit}
    \end{subfigure}
    \begin{subfigure}[b]{0.64\textwidth}
        \setlength{\tabcolsep}{4pt}
        \renewcommand{\arraystretch}{.4}
        \begin{tabular}{l|cccccccccccccccc}
            $\sigma_a$  & $\mathds{1}$ & $\mathds{1}$ & $\mathds{1}$ & $\mathds{1}$ & $\sigma_x$ & $\sigma_x$ & $\sigma_x$ & $\sigma_x$ & $\sigma_y$ & $\sigma_y$ & $\sigma_y$ & $\sigma_y$ & $\sigma_z$ & $\sigma_z$ & $\sigma_z$ & $\sigma_z$\\
            $\sigma_b$  & $\mathds{1}$ & $\sigma_x$ & $\sigma_y$ & $\sigma_z$ & $\mathds{1}$ & $\sigma_x$ & $\sigma_y$ & $\sigma_z$ & $\mathds{1}$ & $\sigma_x$ & $\sigma_y$ & $\sigma_z$ & $\mathds{1}$ & $\sigma_x$ & $\sigma_y$ & $\sigma_z$\\
            $\sigma_c$  & $\mathds{1}$ & $\mathds{1}$ & $\sigma_z$ & $\sigma_z$ & $\sigma_x$ & $\sigma_x$ & $\sigma_y$ & $\sigma_y$ & $\sigma_y$ & $\sigma_y$ & $\sigma_x$ & $\sigma_x$ & $\sigma_z$ & $\sigma_z$ & $\mathds{1}$ & $\mathds{1}$\\
            $\sigma_d$  & $\mathds{1}$ & $\sigma_x$ & $\sigma_y$ & $\sigma_z$ & $\sigma_x$ & $\mathds{1}$ & $\sigma_z$ & $\sigma_y$ & $\sigma_x$ & $\mathds{1}$ & $\sigma_z$ & $\sigma_y$ & $\mathds{1}$ & $\sigma_x$ & $\sigma_y$ & $\sigma_z$\\
        \end{tabular}
        \vspace{2\baselineskip}
        \caption{Valid combinations for Pauli-twirling of the CX gate.}
    \end{subfigure}
    \caption{Pauli-twirling and noise amplification.}
    \label{fig:Paulitwirling}
\end{figure}

\subsection{Pauli-twirling}\label{sec:Paulitwirling}
Before we apply the noise amplification, we convert the non-stochastic errors of CX gates into stochastic errors, see e.g. \cite[section VII]{Li2017} for a detailed description.
One way to achieve this is to apply Pauli-twirling.
Given a finite group $G$ of quantum operations and a quantum channel $\Lambda$ the average
\begin{equation}
    \frac{1}{|G|} \sum_{U\in G} U^\dagger \Lambda U,
\end{equation}
is called a \textit{twirl} of the channel $\Lambda$.
In our case gates $\sigma^a, \sigma^b, \sigma^c, \sigma^d$ are inserted before and after each CX gate $\Lambda$, where $\sigma^i$ is chosen from the twirling set consisting of the Pauli gates $\{ \mathds{1}, \sigma^x, \sigma^y, \sigma^z \}$.
After randomly (with uniform probability) choosing $\sigma^a, \sigma^b$ the gates $\sigma^c,\sigma^d$ are then chosen to satisfy
\begin{equation}
    \sigma ^c \otimes \sigma ^d = \mathrm{e}^{i \theta} \Lambda (\sigma^a \otimes \sigma^b) \Lambda^\dagger.
\end{equation}
This ensures that the overall effect results only in a phase change, which does not change the measurement outcome.
The circuit constructed with Pauli-twirling applied to all CX gates is therefore equivalent to the original circuit.
Figure~\ref{fig:Paulitwirling} shows a schematic depiction of Pauli-twirling as well as all valid combinations for the CX gate.
In practice this method is applicable, if the assumption holds that the qualities of single-qubit gates are an order of magnitude smaller than two-qubit gates.
Twirling should then only have a negligible effect on the fidelity of the expectation value on NISQ devices.
Figure~\ref{fig:Expectationvalues} indicates that noise manifests itself in an increase of the variance of the distribution. There is no effect for the ideal simulator.

\begin{figure}
\centering
    \begin{subfigure}[b]{0.32\textwidth}
        \captionsetup{width=1.2\linewidth}
        \includegraphics[width=\textwidth]{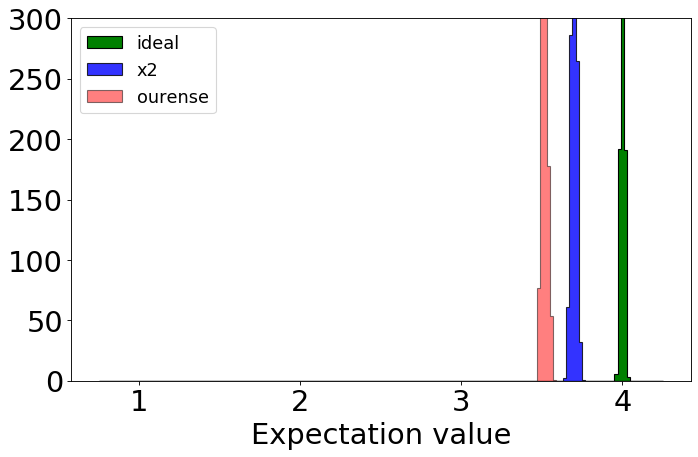}
        \caption{Result without noise amplification $r=1$.}
    \end{subfigure}
    \begin{subfigure}[b]{0.32\textwidth}
        \includegraphics[width=\textwidth]{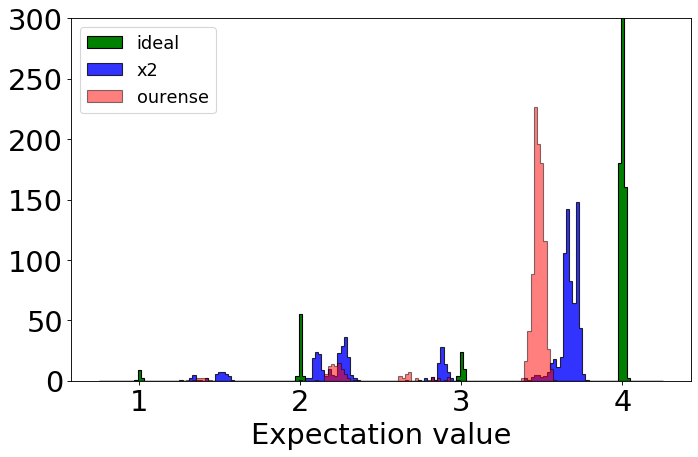}
        \caption{Result for $r=4$.}
    \end{subfigure}
    \begin{subfigure}[b]{0.32\textwidth}
        \includegraphics[width=\textwidth]{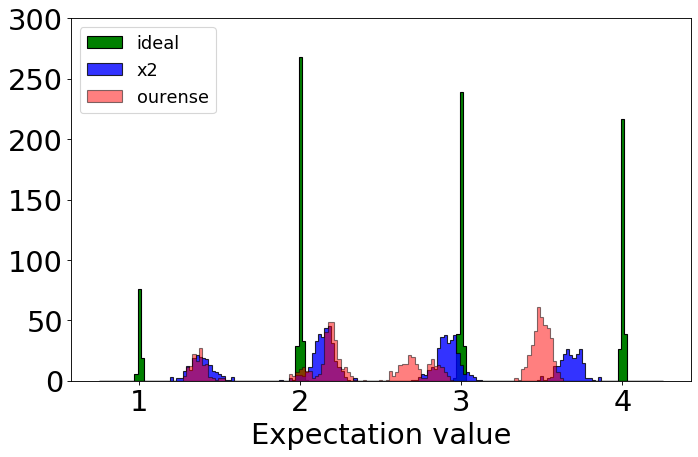}
        \caption{Result for $r=32$.}
    \end{subfigure}
    \begin{subfigure}[b]{0.49\textwidth}
    \centering
        \includegraphics[width=0.75\textwidth]{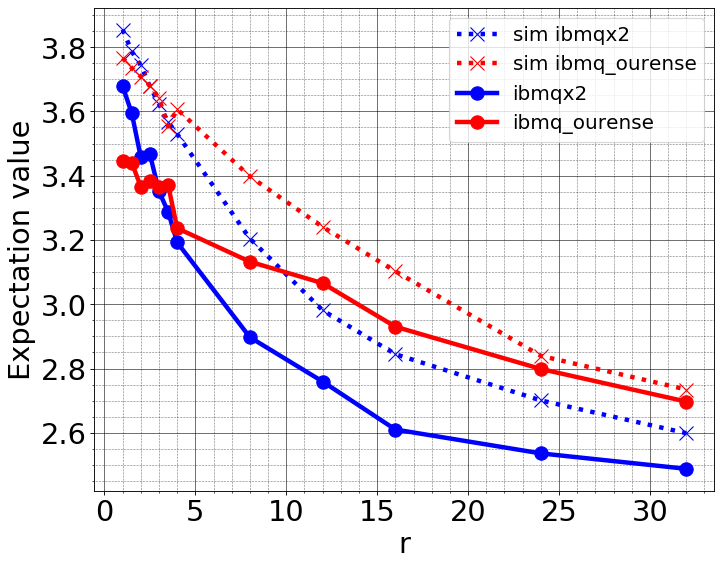}
        \caption{Expectation value obtained for different noise amplification factors $r$.
        }
        \label{fig:rdependency}
    \end{subfigure}
    \begin{subfigure}[b]{0.49\textwidth}
    \centering
     \includegraphics[width=0.75\textwidth]{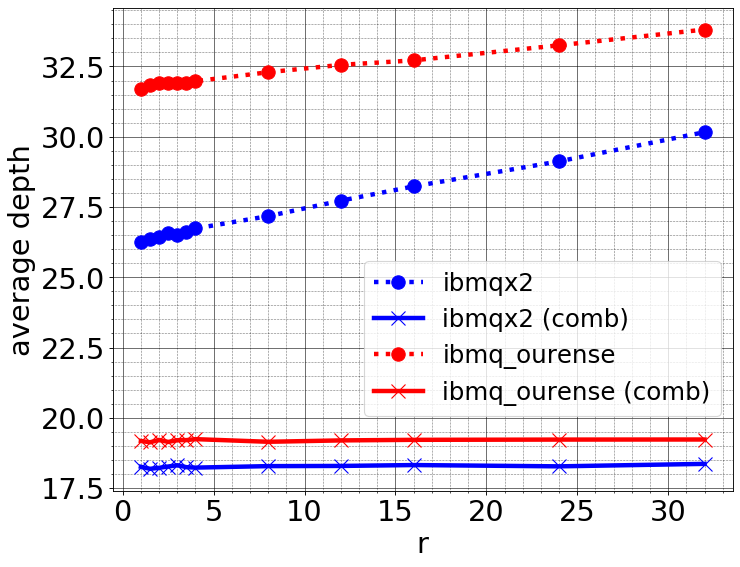}
        \caption{Depth of the circuits for varying $r$. 
        The average depth is roughly constant when combining single qubit gates.
        }
         \label{fig:circuitdepth}
    \end{subfigure}
    \caption{
    Effect of artificially amplifying the noise on Max's circuit.
    Each sample point uses $N=1024$ randomly generated circuits with 8192 shots per circuit.
    The $r$-dependence of the simulated noise model is much smoother than that of IBM's QX2 device.
    Amplifying the noise with random Pauli gates leads to a multi-peaked distribution, because the expectation value is a sum of different outcomes.
    }
    \protect 
    \label{fig:noiseamplification}
\end{figure}
\subsection{Noise amplification}
In order to amplify the strength of the noise, we will apply random Pauli gates with a probability proportional to the error rate of the CX gate between a given pair of qubits. More precisely this is means applying gates $\sigma^e, \sigma^f$ randomly chosen form the set of Pauli gates $\{ \mathds{1}, \sigma^x, \sigma^y, \sigma^z \}$ after the twirled CX gates with probability $(r-1)\epsilon_{i,j}$, see a depiction in Figure~\ref{fig:pauliandnoisecircuit}.
Note that there are only 15 possible choices for  $\sigma^e \otimes \sigma^f$, since $\mathds{1}\otimes \mathds{1}$ must be excluded because it does not increase the error.
Here, $\epsilon_{i,j}$ is the two-qubit gate error rate between qubits $q_i$ and $q_j$.
On average this increases the error rate to the desired value $\epsilon_\text{new} = \epsilon_{i,j} + (r-1)\epsilon_{i,j} = r\epsilon_{i,j}$.

Figure~\ref{fig:noiseamplification} shows the result for both the simulated error models and the real quantum devices.
The assumption that the expectation value of an observable depends smoothly on $r$ seems to hold for the simulator with the IBM QX2 noise model and the IBM QX2 device, but less so for the IBM ourense device, see Figure~\ref{fig:rdependency}.
This is likely because some of the underlying assumptions of the method are violated for the ourense device, e.g., the existence of non-Markovian noise, spatially or temporally correlated noise, etc.
The result shown in Figure~\ref{fig:rdependency} seems to justify the assumption of the exponential variant of the extrapolation method presented in \cite{Endo2018}.

Additional insight is provided by looking at the distribution for $r\in\{1,4,32\}$.
Since we increase the noise of CX gates artificially by adding Pauli gates, this means that other outcome strings are becoming more likely.
Figure~\ref{fig:noiseamplification}a-c shows that expectation values of 1, 2, 3 become increasingly likely.
The result is a multi-peaked distribution.
The noisy results show the same basic behaviour as the ideal circuit.
In general, the noise models seem to lead to better estimates of the expectation values than the real quantum devices, limiting their usefulness somewhat.

\newpage
\subsection{Error mitigation of measurement noise}
\begin{figure}
    \centering
    \begin{subfigure}[b]{0.49\textwidth}
        \centering
        \includegraphics[width=.75\textwidth]{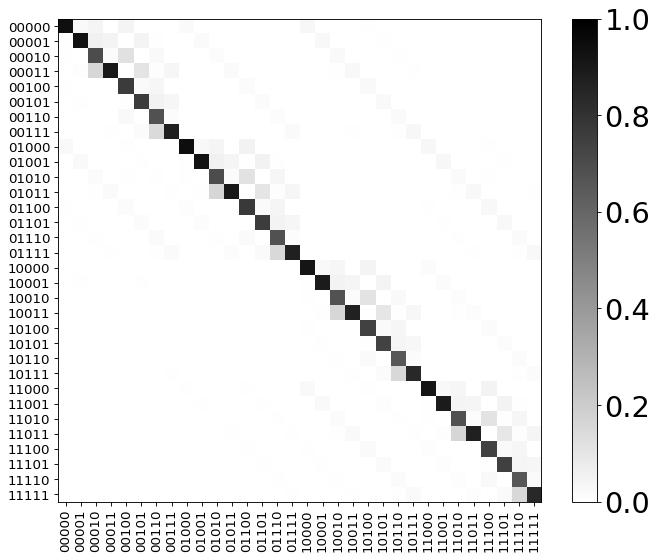}
        \caption{Conditional probabilities $P(\ket{i}|\ket{j})$ for IBM's QX2 device.}
        \label{fig:measfilt}
    \end{subfigure}
    \begin{subfigure}[b]{0.49\textwidth}
        \centering
        \includegraphics[width=1\textwidth]{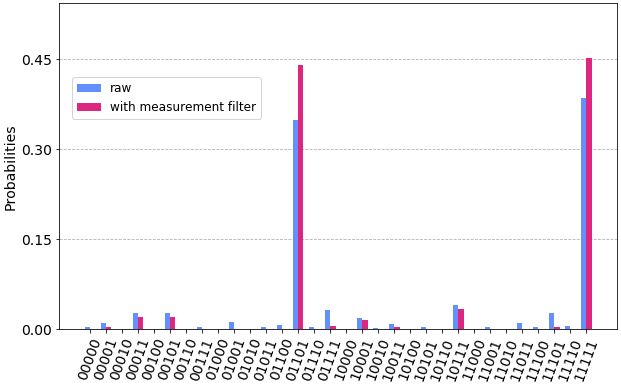}
        \caption{Effect of measurement filter.}
        \label{fig:effectmeasfilt}
    \end{subfigure}
    \caption{The results from the prepared vs measured state can be used to construct a filter to mitigate measurement errors. The filter applied to Max's circuit improves the probabilities.}
    \label{fig:Pmeasnoise}
\end{figure}

Measurement noise is another major source of error.
Here, we use the model that assumes spatially uncorrelated errors of a bit flip.
We compute the probability that the state $\ket{i}$ is observed if the state $\ket{j}$ is prepared, i.e. the conditional probability $P_{i,j}\coloneqq P(\ket{i}|\ket{j})$.
The matrix
\begin{equation}
    P=\left[\begin{matrix}
P_{1,1}&\dots&P_{1,2^n}\\
\vdots&\ddots&\vdots\\
P_{2^n,1}&\dots&P_{2^n,2^n}\\
\end{matrix}\right],
\end{equation}
is a (right) stochastic matrix, as $\sum_j P_{i,j} = 1$.
In the absence of noise $P_{i,j}=\delta_{i,j}$, but measurement (and other) noise leads to non-zero off-diagonal entries.
The resulting probabilities for IBM's QX2 are shown in Figure~\ref{fig:measfilt}.

Let us assume that, for a quantum computer, we are given $P$ and a probability distribution $D_\text{noisy}$ induced by measurement of a quantum state $\ket{\Psi}$. Using the equation
\begin{equation}\label{eq:measfil}
    D_\text{noisy} = P D_\text{ideal},
\end{equation}
we can retrieve the ideal distribution $D_\text{ideal}$ of $\ket{\Psi}$.
As an example, we are preparing the Bell state $\ket{\Psi}=1/\sqrt{2}(\ket{00}+\ket{11})$, but the resulting distribution is $D_\text{noisy}=(13,2,2,13)/30$. In addition we have that  $P_{i,j}$ is $0.8$ if $i=j$ and $0.2/3$ otherwise. By solving \eqref{eq:measfil} we can then retrieve the noiseless distribution $D_\text{ideal}=(1/2,0,0,1/2)$.

In order for the method to work, measurement errors must be at least one order of magnitude larger than state preparation and the execution of the $X$ gate.
This condition is satisfied for IBM's QX2 and ourense device, compare Figure~\ref{fig:realdevices}.
In addition it must be mentioned that this requires an exponential amount (in the number of qubits) of states to be prepared and measured to build the matrix $P$.
In this work we use the implementation provided by Qiskit\cite{Qiskit}.
Figure~\ref{fig:effectmeasfilt} and the column for $E_1$ in Table~\ref{tab:relerror} show a clear improvement by applying the measurement filter for Max's circuit.

\subsection{Overall results}\label{sec:results}
\begin{figure}
    \centering
    \begin{subfigure}[b]{0.46\textwidth}
        \centering
        \includegraphics[width=\textwidth]{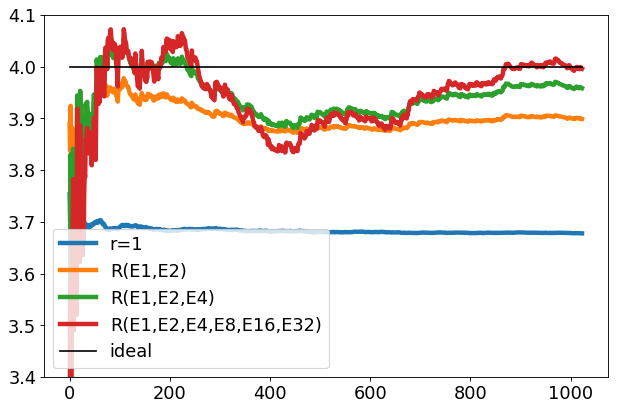}
        \caption{Convergence of Expectation values for IBM QX2.}
        \label{fig:convqx2}
    \end{subfigure}
    \begin{subfigure}[b]{0.46\textwidth}
        \centering
        \includegraphics[width=\textwidth]{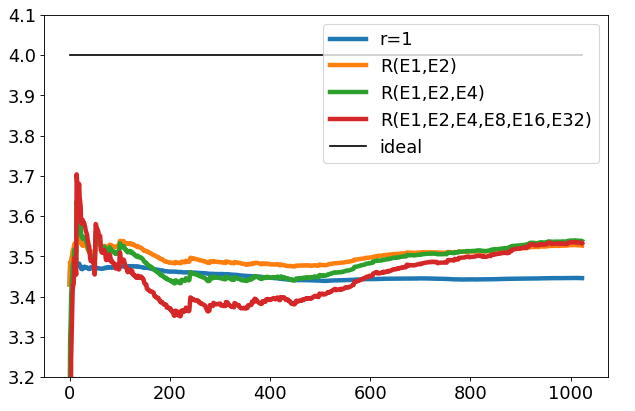}
        \caption{Convergence of Expectation values for IBM ourense.}
        \label{fig:convourense}
    \end{subfigure}
    \begin{subfigure}[b]{1\textwidth}
    \centering
        \setlength{\tabcolsep}{2pt}
        \renewcommand{\arraystretch}{.2}
        \begin{tabular}{lcccccc}
        &&&&&&\\
        &&&&&&\\
        &$E_1$ & $R(E_1,E_2)$ & $R(E_1,E_2,E_4)$ & $R(E_1,E_2,E_4,E_8)$ & $R(E_1,E_2,E_4,E_8,E_{16})$ & $R(E_1,E_2,E_4,E_8,E_{16},E_{32})$  \\
        &&&&&&\\
        \hline
        &&&&&&\\
sim X2&3.85&3.97&3.97&3.97&3.97&3.96\\
&&&&&&\\
sim ourense&3.77&3.82&3.83&3.83&3.83&3.83\\
&&&&&&\\
X2&3.68&3.90&3.96&3.98&3.99&4.00\\
&&&&&&\\
ourense&3.45&3.53&3.54&3.54&3.53&3.53\\
&&&&&&\\
\hline
&&&&&&\\
sim X2 (mf) &3.95&4.06&4.07&4.06&4.06&4.06\\
&&&&&&\\
sim ourense (mf) &3.87&3.93&3.94&3.94&3.94&3.94\\
&&&&&&\\
X2 (mf) &3.80&4.04&4.10&4.12&4.13&4.14\\
&&&&&&\\
ourense (mf) &3.62&3.71&3.72&3.72&3.71&3.71\\
        \end{tabular}
    \caption{Expectation values achieved by simulators and real quantum devices. "mf" denotes results using the measurement filter.}
    \label{tab:relerror}
    \end{subfigure}
    \caption{Results achieved using the simulator and IBM's quantum devices. The symbols $E_1, E_2, \cdots E_{32}$ denote the expectation value for the noise amplification factor $r$ equal to $1, 2, \cdots, 32$ respectively. $R(.)$ denotes the Richardson extrapolation.
    All results were obtained by using 1024 repetitions with 8192 shots each.}
    \label{fig:RichardsonExtrapolation}
\end{figure}

In all of our experiments we generate $N$ circuits randomly with Pauli-twirling and a noise amplification factor $r$.
Each of these circuits is called a "repetition" and uses $8192$ shots.
The number of this random circuits (repetitions) have to be large enough to cover the whole sample space.
Max's circuits has 9 CX operations, which is why we used $N=1024$ repetitions.
We can see in Figures~\ref{fig:RichardsonExtrapolation}\subref{fig:convqx2},\subref{fig:convourense} that this number is sufficient for convergence.
The results for other circuits are similar.

Figure~\ref{fig:RichardsonExtrapolation} shows the convergence of the circuits and the effect of error mitigation techniques on the expectation value.
With $E_r=E(r)$ we denote the expectation value achieved with amplification factor $r$, and by $R(.)$ the Richardson extrapolation.
Without error mitigation the expectation value for the X2 device is closer to the theoretical value than the ourense device. The execution on real devices leads to a worse result as the simulators. Compare also Figure~\ref{fig:distributionwithpauli}

Applying Richardson extrapolation clearly improves the resulting expectation value in all cases.
With increasing number of terms the achieved estimate of the expectation value seems to converge.
For the X2 device, already 2 to 3 terms are sufficient to achieve a very good approximation.
For the ourense device, the results are not as good.
This is most likely due to longer circuit depth and higher measurement errors.

Applying the measurement error filter alone helps to improve the expectation value as well, particularly when using a quantum simulator.
However, on the real devices, the results are not as good as for the Richardson extrapolation.
Combining Richardson extrapolation and measurement error filter seems to only work for the ourense model and device.

\section{Relationship to benchmarking}
We would like to remark that circuits from the proposed game could be used to benchmark quantum devices.
In randomized benchmarking, see e.g., \cite{knill2008randomized}, random sequences are generated from the Clifford group, including a computed reversal element such that the overall unitary is the identity up to a global phase factor.
Quantum Poker has a random part, the "community cards", and a part that depends on the strategy of the individual player, see Equation~\eqref{eq:QP}.
In order to use quantum poker in this setting, one would need to make the following changes:
\begin{itemize}
    \item Negate the original definition of winning to be maximizing the number of 0's that are measured.
    \item Provide a player with enough cards, such that the overall unitary of an identity modulo global phase can be created, i.e., such that $U=e^{i\theta}V$.
\end{itemize}
With this changes the game can be used to benchmark subspaces of 5 qubits.
A generalization to $n$ qubits is also straight forward.

\section{Availability of Data and Code}
The open source python/jupyter notebook implementation of the game is available at \url{https://github.com/sintefmath/quantumpoker} and the complete code for reproducing the results obtained in this article is available at \url{https://github.com/OpenQuantumComputing}.

\section{Conclusion}\label{sec:conclusion}
We have presented a game intended to serve as a pedagogical tool for learning the basic rules of quantum computers.
The aim was to make it a fun experience in order to get more people acquainted with the rules of quantum computing and raise interest in algorithms and error mitigation.
It could therefore help to avoid a shortage in experts when quantum computers become commercialized.
To make the threshold for acquiring the game lower, it could be made available on smartphones as well.
In order to make the game more difficult one could generate a completely random initial state.
The quantum Poker game can be easily simulated on classical computers because it requires only 5 qubits. However, on contemporary quantum computers the use of multiple error-prone CX gates and measurement operations, gives a large error in the output state of the circuits.
We have presented and discussed several error mitigation techniques that produce better estimates of expectation values.
The overall results rely on Pauli-twirling to convert non-stochastic errors of CX gates into stochastic errors.

\bibliographystyle{plainnat}
\bibliography{references}

\end{document}